\documentclass[a4paper,11pt]{article}
\usepackage{postp}

\def\EPOS{{\sc Epos}}
\newcommand{\QGSJETIII}{\textsc{QGSJet-III}}

\newcommand{\QGSJETIId}{\textsc{QGSJetII.04}}
\newcommand{\EPOSLHC}{\textsc{Epos~LHC}}
\newcommand{\EPOSLHCR}{\textsc{Epos.LHC-R}}

\newcommand{\EPOSFour}{\textsc{Epos4}}
\newcommand{\SIBYLL}{\textsc{Sibyll~2.3}e}
\newcommand{\URQMD}{\textsc{UrQMD~3.4}}

\newcommand{\xmax}{X_\text{max}}

\title{EPOS.LHC-R : a global approach to solve the muon puzzle}
 \ShortTitle{EPOS.LHC-R}

\author[a]{Tanguy Pierog}
\author[b]{Klaus Werner}

\affiliation[a]{Karlsruhe Institute of Technology (KIT), Institute for Astroparticle Physics, \\
  Postfach 3640, D-76021 Karlsruhe, Germany}

\affiliation[b]{SUBATECH, University of Nantes -- IN2P3/CNRS -- ,\\
IMT Atlantique, Nantes, France}

\emailAdd{tanguy.pierog@kit.edu}

\abstract{The hadron production in the simulation of extensive air showers is a long standing problem and the origin of large uncertainties in the reconstruction of the mass of the high energy primary cosmic rays. Hadronic interaction models re-tuned after early LHC data give more consistent results among each other compared to the first generation of models, but still can't reproduce extended air shower data (EAS) consistently resulting in the so-called "muon puzzle". Using more recent LHC data like in the QGSJET-III model improve further the description of EAS by such a model but is not enough to resolve the discrepancy. On the other hand, the EPOS project is a theoretical global approach aiming at describing data from very fundamental electron-positron interactions to central heavy ions collisions. We will demonstrate that this approach can provide new constraints, changing the correlation between the measured data at mid-rapidity and the predicted particle production at large rapidities, which drive the EAS development. Thus, using the same accelerator data, different predictions are obtained in air shower simulations in much better agreement with the current air shower data (for both the maximum shower development depth Xmax and the energy spectrum of the muons at ground). Using the EPOS LHC-R model, the detailed changes will be addressed and their consequences on EAS observable at various energies.}

\begin{document}
\maketitle

\section{Introduction}
\label{intro}

Despite all the efforts made to take into account the first results of proton-proton collisions at the LHC in hadronic interaction models used for air-shower simulations, the observed number of muons, their height of production, or even the depth of shower maximum are still not reproduced consistently by the models~\cite{Pierog:2017awp,PierreAuger:2024neu}. Furthermore, the differences among model predictions introduce uncertainties in cosmic-ray data analysis, which are currently smaller than in the past but still exceed the experimental uncertainties in certain cases. Nevertheless before claiming the need for ``new physics'', it is important to guarantee that all the standard QCD physics is properly taken into account in these models. For that, it is necessary to go beyond the simplest observables which are usually used to test them. After almost fifteen years of running, the various LHC experiments provided a large amount of complex data to analyze and understand, in particular, thanks to the correlations between different observables, which are not yet fully investigated.

Among the hadronic interaction models used for air-shower analysis, only \EPOSLHC~\cite{Pierog:2013ria} includes all the features needed to have a detailed description of the correlation between various observables~\cite{Pierog:2017awp}. Indeed, the core-corona approach in this model, which allows the production of a collective hadronization phase, appears to be a key element to reproduce LHC data. Before LHC, it was usually accepted that
hydrodynamical phase expansion, for instance due to the formation of a quark-gluon plasma, was possible only in central heavy-ion collisions.
Proton-nucleus (pA) collisions were then used as a reference to probe the effect
of such collective behavior (final state) but with some nuclear effect 
at the initial state level, while proton-proton (pp) interactions were free of
any nuclear effects. With the LHC operated in pp, pPb and PbPb mode, it is now possible to 
compare high-multiplicity pp or pPb events with low-multiplicity PbPb events
(which 
correspond to the same number of particles measured at mid-rapidity) and surprisingly, the very same 
phenomena are observed~\cite{ALICE:2016fzo,Pan:2017vrs} concerning the soft-particle production.

At the same time, the results compiled by the Working group on Hadronic Interactions and Shower Physics (WHISP)~\cite{Dembinski:2019uta} clearly indicate that the discrepancy between the muon production in simulations and data gradually increases with energy. It is a strong
indication of a different hadronization than the one used in the current hadronic models~\cite{Baur:2019cpv,Anchordoqui:2022fpn}, including \EPOSLHC, which does not have enough core contribution according to data~\cite{ALICE:2016fzo} published after the release of the model in 2012.

But before any claim based on the core-corona model, it is important to fully understand the corona part (based on string fragmentation) and check all possible sources of uncertainty which could lead to a change in muon production. In Section~\ref{sec:rho}, we will check the uncertainty on the $\rho$ resonance production in data and its impact on the muon production. 

Furthermore, other studies showed some deficiencies of \EPOSLHC\ and other models not only for the number of muons but also for $\xmax$ measurements~\cite{PierreAuger:2024neu}. Since this is the most important observable for the mass composition of cosmic rays, one should check the model against the latest LHC data in terms of elasticity and multiplicity distributions which are key factors for the shower development~\cite{Ulrich:2010rg}. This will be done in Section~\ref{sec:xmax} using the final version of the new tune of \EPOS, called \EPOSLHCR, in comparison to the latest version of the other models used for air shower simulation, namely \QGSJETIII~\cite{Ostapchenko:2024myl,Ostapchenko:2024myl} and \SIBYLL~\cite{Riehn:2019jet}. Finally, a summary is given in Section~\ref{sec:summary}.

\section{Hadronization and air-shower physics}
\label{sec:rho}

The dominant mechanism for the production of muons in air showers is
via the decay of light charged mesons. The vast majority of mesons are
produced at the end of the hadron cascade, after typically five to ten
generations of hadronic interactions (depending on the energy and
zenith angle of the cosmic ray).
The energy carried by neutral pions, however, is directly fed into the
electromagnetic shower component and is not available for further
production of more mesons and subsequently muons. Thus, the energy carried
by hadrons other than neutral pions is typically available to produce
more hadrons and ultimately muons in following interactions and
decays. As explained in Ref.~\cite{Baur:2019cpv,Pierog:2006qv}, the ratio of the average electromagnetic to average hadronic
energy, called $R$, and its dependence on
center-of-mass energy, are thus related to the muon
abundance in air showers: if this energy ratio is smaller (larger),
more (less) energy is available for the production of muons at the end
of the hadronic cascade and ultimately more (fewer) muons are produced.

\begin{figure*} 
  \centering
  \includegraphics[width=0.9\textwidth,angle=0]{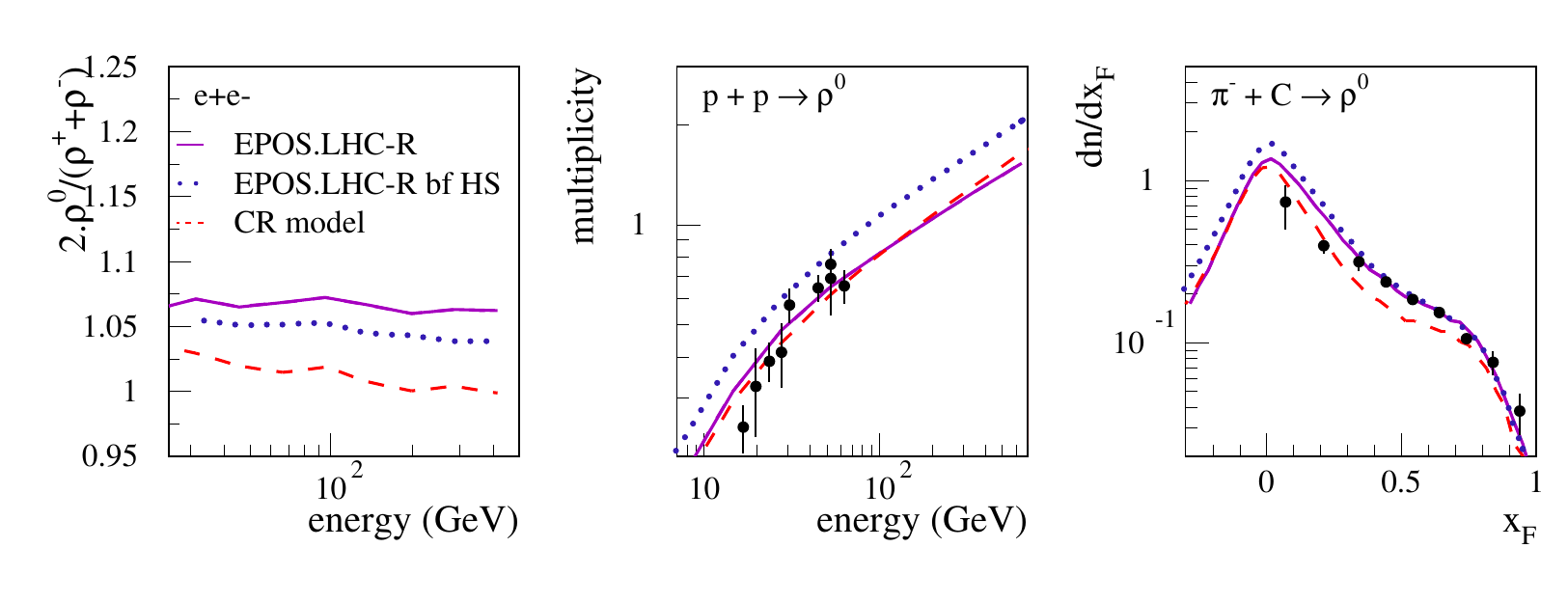}
\vspace{-4ex}
  \caption{ Left-hand side: $\rho^0/\rho^\pm$ in e$^+$e$^-$ interactions as a function of the center-of-mass energy. Middle: $\rho^0$ production in $p+p$ interactions as a function of the center-of-mass energy. Right-hand side: $\rho^0$ yield as measured by the NA61 collaboration at 158 GeV/c in $\pi^-+C$ interactions \protect\cite{NA61SHINE:2017vqs} as a function of the Feynman momentum fraction. Simulations are given for \EPOSLHCR\ before hadronic scattering effect (``bf HS'',dotted line), \EPOSLHCR\ including HS (full line), and a generic ``CR model'' tuned without $\eta\prime$ and $f_0$ (dashed line). }
  \label{fig:rho}
\end{figure*}

In a simplified model where the only existing hadrons would be pions and considering perfect isospin symmetry, the ratio $R$ would be fixed to 0.5. It was already shown~\cite{Pierog:2006qv} that when more particles species are considered (baryons, kaons) this value is reduced and then depends on the hadronization type~\cite{Baur:2019cpv} (string fragmentation in corona or statistical decay in core). But since the pions are dominating, it is difficult to change $R$ a lot. In fact, the pions them-selves can be either produced directly or via the decay of higher mass resonances such as $\rho$ mesons. With perfect isospin symmetry, where the same number of $\rho^0$, $\rho^+$ and $\rho^-$ are produced, the same fractions 1:1:1 are obtained for the pions. Using this theoretical constrain, hadron production from e$^+$e$^-$ collisions can be reproduced within the uncertainties.
In~\cite{Riehn:2023} it is shown that an ad-hoc replacement of $\pi^0$ by $\rho^0$ has a strong influence on muon production because $\rho^0$ decays into charged pions changing significantly the value of $R$. But this would break the isospin symmetry, which is a very basic conservation law. In fact in \EPOSLHC, isospin symmetry was slightly broken (different effective mass for u and d quarks) to better reproduce e$^+$e$^-$ data using a string fragmentation where only  $\pi^0$, $\rho^0$, $\omega$ and $\eta$ were produced based on a pair of $u-\bar{u}$, $d-\bar{d}$, (or $s-\bar{s}$ for $\eta$). This limited number of resonances is common to all cosmic ray (CR) models all lead to the ratio between neutral $\rho$ and charged $\rho$ shown on Fig.~\ref{fig:rho} left-hand side as a dashed line. It is close to 1 but not exactly one because of the decay of other higher mass resonances.
In \EPOSLHCR, the isospin is now perfectly conserved in the fragmentation process, but the two high mass resonances $\eta\prime$ and $f_0$ have been added to the list of neutral particles. The rest of the e$^+$e$^-$ is equally well reproduced in both cases, but now the $\eta\prime$ resonance will decay into $\rho^0$ but not in $\rho^\pm$ breaking the expected isospin symmetry. The result can be seen as the dotted line on Fig.~\ref{fig:rho} left-hand side, 5\% above the ``CR model'' dashed line.

In addition, \EPOSLHCR\ is based on the idea of a ``global approach'' developed for \EPOSFour~\cite{Werner:2023zvo,Werner:2023jps}, in which all physics processes are taken into account (if they apply) for all systems from e$^+$e$^-$ to PbPb. In particular, the hadronic scattering (HS) which is important to reproduce PbPb data is also applied in e$^+$e$^-$ or pp, slightly changing the results in particular for heavy resonances like $\rho$s. This effect, simulated with \URQMD~\cite{Bass:1998ca,Bleicher:1999xi}, reduce the final number of $\rho$s leading to different tuning parameters and modify the $\rho^0/\rho^\pm$ ratio as shown by the full line on Fig.~\ref{fig:rho} left-hand side. The effect can be seen in the direct measurement of $\rho^0$ in low energy $p+p$ interaction where we see on then central Fig.~\ref{fig:rho}, that to reproduce the data with HS, the model should first produce a larger yield which is then reduced by HS. This effect is particularly interesting because it depends on the phase space. It plays a role only when the particles are slow and close enough to each other to interact after their initial production. This is only possible at mid-rapidity while at large rapidities the particles are too fast to suffer such effect. As a consequence it allows to better reproduce the recent NA61 data~\cite{NA61SHINE:2017vqs} with a large fraction of forward $\rho^0$ without having too many at mid-rapidity as shown on Fig.~\ref{fig:rho} right-hand side, while the standard ``CR model'' type cannot reproduce both phase space at the same time with the same parameters for the string fragmentation.

Having about 7\% more $\rho^0$ without changing charged $\rho$s is decreasing R by about 5\% and as a consequence the number of muons is increased by about 7\% compared to other model as shown on Fig.~\ref{fig:xmax} left-hand side. Additionally the baryon production from beam remnant has been reduced compared to \EPOSLHC, to better reproduce NA61 data~\cite{NA61SHINE:2017vqs}, this should lead to a reduction of the muon production. But since the core-corona model is also applied at lower multiplicity according to~\cite{ALICE:2016fzo}, it somewhat compensate this effect.

\section{Multiplicity and elasticity at LHC}
\label{sec:xmax}

Using the simple Heitler-Matthews model~\cite{Matthews:2005sd} or studying Monte-Carlo simulations~\cite{Ulrich:2010rg}, it is well known that the inelastic cross-section, the elasticity and the multiplicity of hadronic interactions, together with the mass and energy of the primary cosmic rays are the most important factors which determine the position of the maximum air shower development $\xmax$. For \EPOSLHC, this was fixed using the early data of different LHC experiments at 7~TeV center-of-mass energy. Since then, new and more precise data have been released at 13~TeV or using lead as a projectile giving new constraints to the models which should be taken into account to give proper $\xmax$ predictions together with other bug fix on nuclear fragmentation as described in~\cite{Pierog:2023ahq}

The elasticity is important for the shower development but is complicated to measure at LHC. The most direct measurement is by the LHCf collaboration~\cite{LHCf:2020hjf} and can be used to constrain the models in particular once the pion exchange process is taken into account. It leads to larger elasticity in both  \EPOSLHCR\ and \QGSJETIII.

Some other distributions could be sensitive to the energy fraction of the leading particle. The average multiplicity of pp and pPb interactions, another important measurement for $\xmax$, is well described by the hadronic models. In the meantime, the fluctuations of the multiplicity in pPb collision have been published and is, in fact, not so well reproduced by the two models which can run with lead projectile \EPOSLHC\ and \QGSJETIII. In fact, the latter is linked to the elasticity because the frequency of low multiplicity events is related to highly elastic events while the high multiplicity events are linked to low elasticity. To better describe the fluctuations of the multiplicity~\cite{CMS:2013jlh} in particular with nuclear effects such as the core-corona, the color transparency of the beam particles has been introduced in \EPOSLHCR. This has a direct impact on the elasticity prediction.

\begin{figure*}[b]
  \centering
  \includegraphics[width=0.9\textwidth,angle=0]{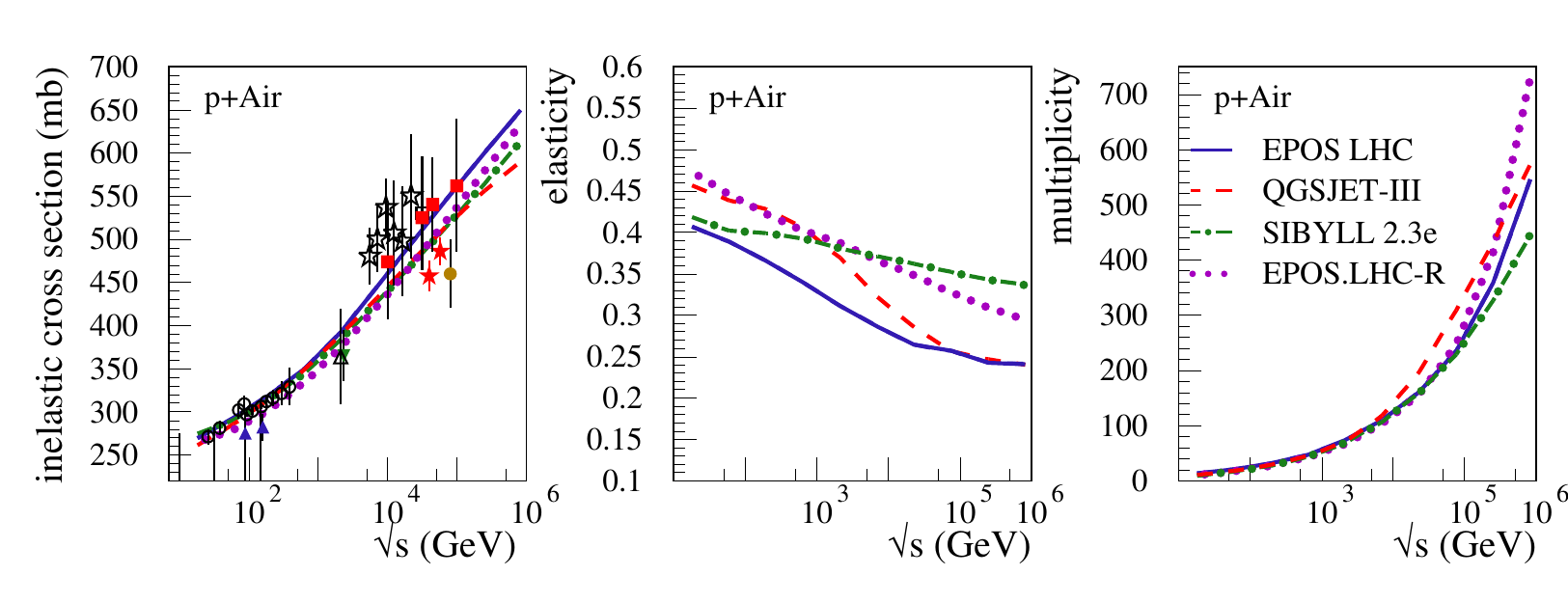}
\vspace{-2ex}
  \caption{Inelastic cross-section (left-hand side), elasticity (energy fraction of the leading particle) (middle) and multiplicity (right-hand side)  for p-air interactions as a function of center of mass energy. Simulations are done with \EPOSLHCR\ (dotted line), \EPOSLHC\ (full line), \QGSJETIII\ (dashed line), and \SIBYLL\ (dash-dotted line). Points are data from~\protect\cite{PDG98}.}
  \label{fig:air}
\end{figure*}

\subsection{Hadronic interactions in Air and $\xmax$}

In order to understand the impact of these modifications on the $\xmax$ predictions by \EPOSLHCR\ we can first check the p-air cross-section and elasticity. In Fig.~\ref{fig:air}, the inelastic p-air cross-section is shown on the left-hand side, the elasticity on the middle panel and the multiplicity is on the right-hand side. The lines are for \EPOSLHCR\ (dotted line), \EPOSLHC\ (full line), \QGSJETIII\ (dashed line), and \SIBYLL\ (dash-dotted line), and the points are data from~\protect\cite{PDG98}. \EPOSLHCR\ has a cross-section about 5\% lower than \EPOSLHC\ on the full energy range, an elasticity which is increased by around 20\%, and a multiplicity increased by about 30\% at the highest energy only.

\begin{figure*}[hptb]
\centering
\includegraphics [width=0.45\textwidth]{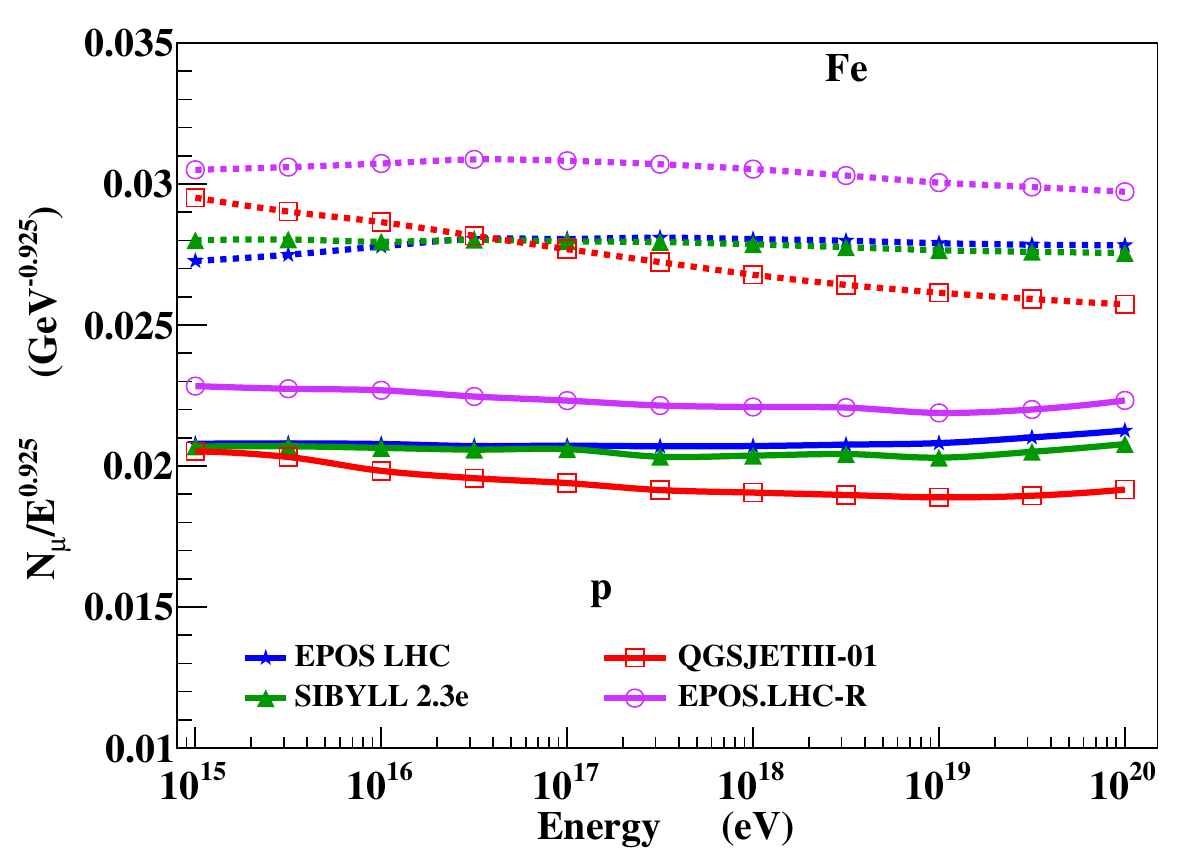}
\includegraphics [width=0.45\textwidth]{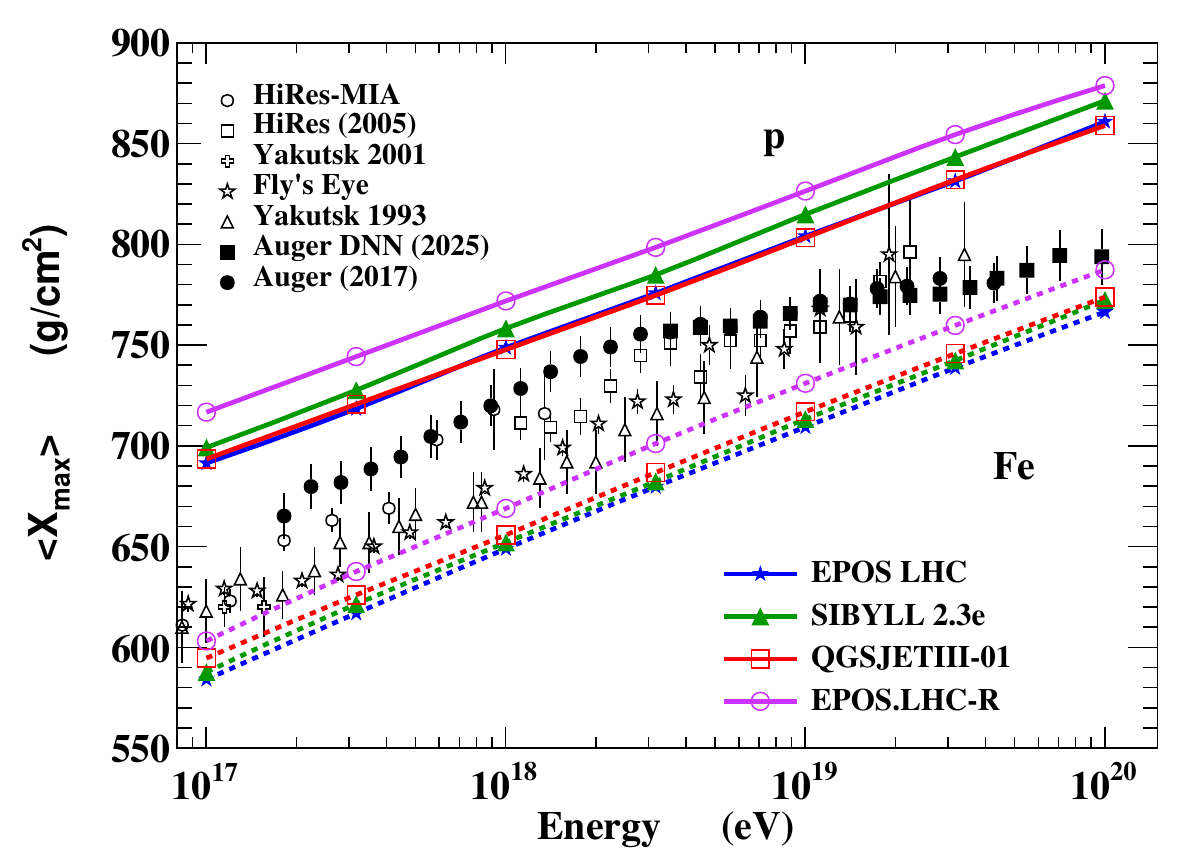}
\caption{\label{fig:xmax} Number of muons normalized by $E^{0.925}$ (left-hand side) and $\langle X_{\rm max} \rangle$ (Right-hand side) for proton- and iron-induced showers as a 
  function of the primary energy. Predictions of different high-energy 
  hadronic interaction models are presented with full lines for proton and 
  dashed lines for iron 
  with full triangles for \SIBYLL, open circles for
  \EPOSLHCR, open squares for
  \QGSJETIII, full stars for \EPOSLHCR\ with no Isospin symmetry Broken (left-hand side) or
  \EPOSLHC\ (right-hand side). Refs. to the data 
  can be found in \protect\cite{Bluemer:2009zf} and \protect\cite{Aab:2014kda}.}

\end{figure*}

These is quite a big difference for the elasticity, and as shown in Fig.~\ref{fig:xmax} right-hand side, this increases $\xmax$ by about 25~g/cm$^2$ which is larger than the difference between \EPOSLHC\ (line with full stars) and \SIBYLL\ (line with full triangles). So, in fact, \EPOSLHCR\ (line with open circles) predicts larger $\xmax$ values than \SIBYLL, and also deeper than the predictions from the new \QGSJETIII\ (line with open squares it-self 15~g/cm$^2$ deeper than the old \QGSJETIId) and the effect is even larger for iron induced showers.

In terms of mass composition, it means that the mean logarithmic mass deduced from $\xmax$ measurements using \EPOSLHCR\ is now compatible with iron showers at the highest energies. As a consequence it will also reduce the muon deficit in simulation compared to data. Since the muon number is also larger for \EPOSLHCR\ compared to the other models, the correlation between the number of muons and  $\xmax$ is now in better agreement with the Pierre Auger Observatory data then the other models~\cite{PierreAuger:2025}.

\section{Summary}
\label{sec:summary}

The better description of the collective hadronization and, in particular, the hadronic rescattering, which modify the correlation between mid and forward rapidities, has very important consequences for the muon production in air showers. It is also important to first properly define the corona hadronization taking all possible effects into account.

In string fragmentation used for the corona part, the pions can be generated directly or using $\rho$ resonances as intermediate step. If there was a perfect isospin symmetry, the ratio between neutral and charged pions would be the same in both cases and there will be no impact on the muon production. But we demonstrated here that if high mass resonances production is taken into account, a slight  isospin symmetry breaking with up to 7\% more $\rho^0$ than charged $\rho$ is introduced by the decay of such particles. As a consequence, the number of muons is increased by almost 10\% if in addition the indirect effects of the core-corona model on multiplicity is taken into account together with the effect of the hadronic rescattering on the tuning of model parameters.

Furthermore, using updated cross-sections and elasticity from LHC in \EPOSLHCR, $\xmax$ is now 25~g/cm$^2$ deeper, leading to an heavier composition for a given measured $\xmax$.

These results include now all the effect of the core-corona model and hadronic rescattering which are necessary to fully reproduce LHC data, and they demonstrate that all the details of the hadronic interactions are important and should be carefully studied. Simplified models cannot account for all complex hadronization scheme which are observed at the LHC and which can play an important role for the air shower development. This is why a global approach like \EPOS\ is important and should be compared and constrained by a large variety of data. The difference with data in the muon production at high energy has been reduced, but the changes are observed on the full energy scale of CR. So further studies are needed to check the consequences of such predictions at different energies.

\bibliographystyle{JHEP}
\bibliography{Pierog_EPOSLHCR}

\providecommand{\href}[2]{#2}\begingroup\raggedright\begin{thebibliography}{10}

\bibitem{Pierog:2017awp}
T.~Pierog, \emph{{Air Shower Simulation with a New Generation of post-LHC
  Hadronic Interaction Models in CORSIKA}},
  \href{https://doi.org/10.22323/1.301.1100}{\emph{PoS} {\bfseries ICRC2017}
  (2018) 1100}.

\bibitem{PierreAuger:2024neu}
{\scshape Pierre Auger} collaboration, \emph{{Testing hadronic-model
  predictions of depth of maximum of air-shower profiles and ground-particle
  signals using hybrid data of the Pierre Auger Observatory}},
  \href{https://doi.org/10.1103/PhysRevD.109.102001}{\emph{Phys. Rev. D}
  {\bfseries 109} (2024) 102001}
  [\href{https://arxiv.org/abs/2401.10740}{{\ttfamily 2401.10740}}].

\bibitem{Pierog:2013ria}
T.~Pierog, I.~Karpenko, J.M.~Katzy, E.~Yatsenko and K.~Werner, \emph{{EPOS LHC:
  Test of collective hadronization with data measured at the CERN Large Hadron
  Collider}}, \href{https://doi.org/10.1103/PhysRevC.92.034906}{\emph{Phys.
  Rev. C} {\bfseries 92} (2015) 034906}
  [\href{https://arxiv.org/abs/1306.0121}{{\ttfamily 1306.0121}}].

\bibitem{ALICE:2016fzo}
{\scshape ALICE} collaboration, \emph{{Enhanced production of multi-strange
  hadrons in high-multiplicity proton-proton collisions}},
  \href{https://doi.org/10.1038/nphys4111}{\emph{Nature Phys.} {\bfseries 13}
  (2017) 535} [\href{https://arxiv.org/abs/1606.07424}{{\ttfamily
  1606.07424}}].

\bibitem{Pan:2017vrs}
{\scshape ALICE} collaboration, \emph{{Multiplicity and transverse momentum
  evolution of charge-dependent correlations in pp, p-Pb, and Pb-Pb collisions
  at the LHC}}, \href{https://doi.org/10.1088/1742-6596/832/1/012044}{\emph{J.
  Phys. Conf. Ser.} {\bfseries 832} (2017) 012044}.

\bibitem{Dembinski:2019uta}
{\scshape EAS-MSU, IceCube, KASCADE-Grande, NEVOD-DECOR, Pierre Auger, SUGAR,
  Telescope Array, Yakutsk EAS Array} collaboration, \emph{{Report on Tests and
  Measurements of Hadronic Interaction Properties with Air Showers}},  2019
  [\href{https://arxiv.org/abs/1902.08124}{{\ttfamily 1902.08124}}].

\bibitem{Baur:2019cpv}
S.~Baur, H.~Dembinski, M.~Perlin, T.~Pierog, R.~Ulrich and K.~Werner,
  \emph{{Core-corona effect in hadron collisions and muon production in air
  showers}}, \href{https://doi.org/10.1103/PhysRevD.107.094031}{\emph{Phys.
  Rev. D} {\bfseries 107} (2023) 094031}
  [\href{https://arxiv.org/abs/1902.09265}{{\ttfamily 1902.09265}}].

\bibitem{Anchordoqui:2022fpn}
L.A.~Anchordoqui, C.G.~Canal, F.~Kling, S.J.~Sciutto and J.F.~Soriano,
  \emph{{An explanation of the muon puzzle of ultrahigh-energy cosmic rays and
  the role of the Forward Physics Facility for model improvement}},
  \href{https://doi.org/10.1016/j.jheap.2022.03.004}{\emph{JHEAp} {\bfseries
  34} (2022) 19} [\href{https://arxiv.org/abs/2202.03095}{{\ttfamily
  2202.03095}}].

\bibitem{Ulrich:2010rg}
R.~Ulrich, R.~Engel and M.~Unger, \emph{{Hadronic Multiparticle Production at
  Ultra-High Energies and Extensive Air Showers}},
  \href{https://doi.org/10.1103/PhysRevD.83.054026}{\emph{Phys. Rev. D}
  {\bfseries 83} (2011) 054026}
  [\href{https://arxiv.org/abs/1010.4310}{{\ttfamily 1010.4310}}].

\bibitem{Ostapchenko:2024myl}
S.~Ostapchenko, \emph{{QGSJET-III model of high energy hadronic interactions.
  II. Particle production and extensive air shower characteristics}},
  \href{https://doi.org/10.1103/PhysRevD.109.094019}{\emph{Phys. Rev. D}
  {\bfseries 109} (2024) 094019}
  [\href{https://arxiv.org/abs/2403.16106}{{\ttfamily 2403.16106}}].

\bibitem{Riehn:2019jet}
F.~Riehn, R.~Engel, A.~Fedynitch, T.K.~Gaisser and T.~Stanev, \emph{{Hadronic
  interaction model Sibyll 2.3d and extensive air showers}},
  \href{https://doi.org/10.1103/PhysRevD.102.063002}{\emph{Phys. Rev. D}
  {\bfseries 102} (2020) 063002}
  [\href{https://arxiv.org/abs/1912.03300}{{\ttfamily 1912.03300}}].

\bibitem{Pierog:2006qv}
T.~Pierog and K.~Werner, \emph{{Muon Production in Extended Air Shower
  Simulations}},
  \href{https://doi.org/10.1103/PhysRevLett.101.171101}{\emph{Phys. Rev. Lett.}
  {\bfseries 101} (2008) 171101}
  [\href{https://arxiv.org/abs/astro-ph/0611311}{{\ttfamily
  astro-ph/0611311}}].

\bibitem{Riehn:2023}
F.~Riehn et~al., \emph{{Sibyll*: ad-hoc modifications for an improved
  description of muon data in extensive air showers}}, {\emph{PoS} {\bfseries
  ICRC(2023)429} (2023) }.

\bibitem{Werner:2023zvo}
K.~Werner, \emph{{Revealing a deep connection between factorization and
  saturation: New insight into modeling high-energy proton-proton and
  nucleus-nucleus scattering in the EPOS4 framework}},
  \href{https://doi.org/10.1103/PhysRevC.108.064903}{\emph{Phys. Rev. C}
  {\bfseries 108} (2023) 064903}
  [\href{https://arxiv.org/abs/2301.12517}{{\ttfamily 2301.12517}}].

\bibitem{Werner:2023jps}
K.~Werner, \emph{{Core-corona procedure and microcanonical hadronization to
  understand strangeness enhancement in proton-proton and heavy ion collisions
  in the EPOS4 framework}},
  \href{https://doi.org/10.1103/PhysRevC.109.014910}{\emph{Phys. Rev. C}
  {\bfseries 109} (2024) 014910}
  [\href{https://arxiv.org/abs/2306.10277}{{\ttfamily 2306.10277}}].

\bibitem{Bass:1998ca}
S.A.~Bass et~al., \emph{{Microscopic models for ultrarelativistic heavy ion
  collisions}},
  \href{https://doi.org/10.1016/S0146-6410(98)00058-1}{\emph{Prog. Part. Nucl.
  Phys.} {\bfseries 41} (1998) 255}
  [\href{https://arxiv.org/abs/nucl-th/9803035}{{\ttfamily nucl-th/9803035}}].

\bibitem{Bleicher:1999xi}
M.~Bleicher et~al., \emph{{Relativistic hadron hadron collisions in the
  ultrarelativistic quantum molecular dynamics model}},
  \href{https://doi.org/10.1088/0954-3899/25/9/308}{\emph{J. Phys. G}
  {\bfseries 25} (1999) 1859}
  [\href{https://arxiv.org/abs/hep-ph/9909407}{{\ttfamily hep-ph/9909407}}].

\bibitem{NA61SHINE:2017vqs}
{\scshape NA61/SHINE} collaboration, \emph{{Measurement of meson resonance
  production in $\pi ^-+$ C interactions at SPS energies}},
  \href{https://doi.org/10.1140/epjc/s10052-017-5184-z}{\emph{Eur. Phys. J. C}
  {\bfseries 77} (2017) 626}
  [\href{https://arxiv.org/abs/1705.08206}{{\ttfamily 1705.08206}}].

\bibitem{Matthews:2005sd}
J.~Matthews, \emph{{A Heitler model of extensive air showers}},
  \href{https://doi.org/10.1016/j.astropartphys.2004.09.003}{\emph{Astropart.
  Phys.} {\bfseries 22} (2005) 387}.

\bibitem{Pierog:2023ahq}
T.~Pierog and K.~Werner, \emph{{EPOS LHC-R : up-to-date hadronic model for EAS
  simulations}}, \href{https://doi.org/10.22323/1.444.0230}{\emph{PoS}
  {\bfseries ICRC2023} (2023) 230}.

\bibitem{LHCf:2020hjf}
{\scshape LHCf} collaboration, \emph{{Measurement of energy flow, cross section
  and average inelasticity of forward neutrons produced in $ \sqrt{s} $ = 13
  TeV proton-proton collisions with the LHCf Arm2 detector}},
  \href{https://doi.org/10.1007/JHEP07(2020)016}{\emph{JHEP} {\bfseries 07}
  (2020) 016} [\href{https://arxiv.org/abs/2003.02192}{{\ttfamily
  2003.02192}}].

\bibitem{CMS:2013jlh}
{\scshape CMS} collaboration, \emph{{Multiplicity and Transverse Momentum
  Dependence of Two- and Four-Particle Correlations in pPb and PbPb
  Collisions}},
  \href{https://doi.org/10.1016/j.physletb.2013.06.028}{\emph{Phys. Lett. B}
  {\bfseries 724} (2013) 213}
  [\href{https://arxiv.org/abs/1305.0609}{{\ttfamily 1305.0609}}].

\bibitem{PDG98}
{\scshape Particle Data Group} collaboration, \emph{Review of particle
  physics}, {\emph{Eur. Phys. J.} {\bfseries C3} (1998) 1}.

\bibitem{Bluemer:2009zf}
J.~Blumer, R.~Engel and J.R.~Horandel, \emph{{Cosmic Rays from the Knee to the
  Highest Energies}},
  \href{https://doi.org/10.1016/j.ppnp.2009.05.002}{\emph{Prog. Part. Nucl.
  Phys.} {\bfseries 63} (2009) 293}
  [\href{https://arxiv.org/abs/0904.0725}{{\ttfamily 0904.0725}}].

\bibitem{Aab:2014kda}
{\scshape Pierre Auger} collaboration, \emph{{Depth of maximum of air-shower
  profiles at the Pierre Auger Observatory. I. Measurements at energies above
  $10^{17.8}$ eV}},
  \href{https://doi.org/10.1103/PhysRevD.90.122005}{\emph{Phys. Rev. D}
  {\bfseries 90} (2014) 122005}
  [\href{https://arxiv.org/abs/1409.4809}{{\ttfamily 1409.4809}}].

\bibitem{PierreAuger:2025}
{\scshape Pierre Auger} collaboration, \emph{{Update on testing of air-shower
  modelling using combined data of the Pierre Auger Observatory and
  phenomenological consequences}}, {\emph{PoS} {\bfseries ICRC(2025)431} (2025)
  }.

\end{thebibliography}\endgroup

\end{document}